\documentclass[conference]{IEEEtran}
\IEEEoverridecommandlockouts
\usepackage{cite}
\usepackage{amsmath,amssymb,amsfonts}
\usepackage{comment}
\usepackage{graphicx}
\usepackage{mathtools}
\usepackage{textcomp}
\usepackage{xcolor}
\usepackage{setspace}
\usepackage{tabularray}
\usepackage{pifont}
\usepackage{cite}
\usepackage{amsmath,amssymb,amsfonts}
\usepackage{algpseudocode}
\usepackage{graphicx}
\usepackage[final]{microtype}
\usepackage[italic]{mathastext}
\usepackage{libertine}
\usepackage[T1]{fontenc}
\usepackage{textcomp}
\usepackage[varqu,varl]{zi4}
\usepackage[all]{nowidow}
\usepackage{setspace}
\usepackage{colortbl}
\usepackage{url}
\usepackage{hyperref}
\usepackage[hyphenbreaks]{breakurl}

\usepackage{textcomp}
\usepackage{xcolor}
\def\BibTeX{{\rm B\kern-.05em{\sc i\kern-.025em b}\kern-.08em
    T\kern-.1667em\lower.7ex\hbox{E}\kern-.125emX}}

\usepackage[utf8]{inputenc}
\usepackage{kotex}
\usepackage{color}
\usepackage[square,numbers,comma,sort]{natbib}
\usepackage[linesnumbered,ruled,vlined]{algorithm2e}
\usepackage[normalem]{ulem}
    
\usepackage{makecell}
\usepackage{dblfloatfix} 
\usepackage{lipsum}
\usepackage{booktabs}
\usepackage{multirow}
\usepackage{siunitx}
\usepackage{verbatim}

\usepackage{ulem}
\usepackage{url}

\definecolor{mygreen}{RGB}{0,128,0}

\newcommand{\cb}{\textcolor{blue}}
\newcommand{\cred}{\textcolor{red}}

\newcommand{\squishlist}{
\begin{list}{$\bullet$}
	{ \setlength{\itemsep}{0pt}      \setlength{\parsep}{-0pt}
		\setlength{\topsep}{4pt}       \setlength{\partopsep}{0pt}
		\setlength{\listparindent}{-2pt}
		\setlength{\itemindent}{-5pt}
		\setlength{\leftmargin}{1em} \setlength{\labelwidth}{0em}
		\setlength{\labelsep}{0.5em} } }

\newcommand{\squishend}{
\end{list}}

\newcommand{\tbdfixed}{\textsc{DeepVM}}
\newcommand{\tbdplan}{\textsc{DeepPlan}}
\newcommand{\tbdckp}{\textsc{DeepCheck}}

\begin{document}

\title{\huge \tbdfixed{}: Integrating Spot and On-Demand VMs for Cost-Efficient Deep Learning Clusters in the Cloud}

\author{Yoochan Kim$^1$, Kihyun Kim$^1$, Yonghyeon Cho$^{1,\dagger{}}$\thanks{$^{\dagger}$Y. Cho is currently affiliated with LG Electronics.}, Jinwoo Kim$^1$, Awais Khan$^2$, Ki-Dong Kang$^3$\\
Baik-Song An$^3$, Myung-Hoon Cha$^3$, Hong-Yeon Kim$^3$, Youngjae Kim
$^{1,\ddagger{}}$\thanks{$^{\ddagger}$Y. Kim is the corresponding author.}\vspace{0.5pt}\\ \\
\it $^1$Dept. of Computer Science and Engineering, Sogang University, Seoul, Republic of Korea\\
$^2$Oak Ridge National Laboratory, TN, USA, $^3$ETRI, Daejeon, Republic of Korea\\
\rm \small \{vision0913, youkim\}@sogang.ac.kr
}
\maketitle

\setstretch{0.985}

\begin{abstract}

Distributed Deep Learning (DDL), as a paradigm, dictates the use of GPU-based clusters as the optimal infrastructure for training large-scale Deep Neural Networks (DNNs). However, the high cost of such resources makes them inaccessible to many users.
Public cloud services, particularly Spot Virtual Machines (VMs), offer a cost-effective alternative,
but their unpredictable availability poses a significant challenge to the crucial checkpointing process in DDL.
To address this, we introduce \tbdfixed{}, a novel solution that recommends cost-effective cluster configurations by intelligently balancing the use of Spot and On-Demand VMs.
\tbdfixed{} leverages a four-stage process that analyzes instance performance using the FLOPP (FLoating-point Operations Per Price) metric, performs architecture-level analysis with linear programming, and identifies the optimal configuration for the user-specific needs.
Extensive simulations and real-world deployments in the AWS environment demonstrate that \tbdfixed{} consistently outperforms other policies, reducing training costs and overall makespan.
By enabling cost-effective checkpointing with Spot VMs, \tbdfixed{} opens up DDL to a wider range of users and facilitates a more efficient training of complex DNNs.

\end{abstract}



\section{Introduction}
\label{sec:intro}

Distributed Deep Learning (DDL) is 
used to train Deep Neural Networks (DNNs) in environments that involve multiple devices, typically GPUs, or multiple nodes in a network.
By leveraging data parallelism or model parallelism techniques~\cite{eval-multi-level, check-n-run, li2020pytorch,sergeev2018horovod, NEURIPS2019_gpipe,kim2020torchgpipe}, DDL 
addresses the high computational demands of training DNNs and significantly reduces the training. 
However, setting up a GPU-based cluster for large-scale training 
is challenging and costly, making it impractical for small organizations 
to run extensive DDL workloads.
Fortunately, there is an alternative available in the form of public cloud services provided by major providers such as Amazon AWS, Microsoft Azure, and Google Cloud Platform.

Users can run DDL workloads at low costs without the need for an on-premise GPU cluster with public cloud services~\cite{narayanan2020analysis}.
Cloud Service Providers (CSPs) offer various options that cater to several budgetary considerations.
There are \textit{On-Demand VM} instances available at regular prices as well as surplus resources known as \textit{Spot VM} instances, which can be obtained at significantly discounted prices, up to 90\% off (in case of AWS).
Therefore, Spot VM instances have garnered attention among cost-conscious users because they allow for the use of a VM with the same specifications as an On-Demand instance at a much lower price.
CSPs provide Spot VM instances as a specialized resource to maximize profits by utilizing idle resources.
However, it is important to note that there is a trade-off in terms of availability. 
Depending on pricing fluctuations and demand~\cite{CAI201838}, Spot VM instances may be terminated to make the resources available to other users. 
Due to the nature of Spot VM instances provided by the CSP, they are cost-effective but come with the inherent risk of sudden interruption or termination at any time, as noted in SpotLake~\cite{SpotLake}.
To mitigate this risk, users should employ methods such as checkpointing that demonstrate tolerance towards interruptions when using Spot VM instances.

\textbf{Checkpointing Challenges 
in Spot VMs:}
In DDL, Checkpoint-restart is a key technique, widely used for a range of purposes. It is instrumental in preserving the state of the best-performing model, safeguarding against unexpected interruptions, and facilitating debugging during training. 
However, the use of checkpoint-restart in Spot VM environment presents a significant challenge. The main issue arises when a Spot VM instance is preempted. In such cases, users lose access to the VM's storage, which contains the checkpoint data. This loss of access makes it impossible to use the Spot VM's storage as a reliable medium for checkpointing. 
Since checkpointing relies on persistently storing the state of the DNN model and training process, the inability to access the storage on a preempted Spot VM renders the checkpointing process ineffective. {This inherent risk of Spot VMs being preempted and losing storage access severely limits their suitability for checkpoint-restart in DDL, where data integrity and continuity are crucial.}

Thus, other options must be considered for checkpointing on Spot VMs. One can adopt external cloud-based storage solutions. 
However, 
these solutions come with several non-trivial challenges, e.g., 
significant costs associated with data transfer and storage, making them economically unfeasible for frequent checkpoints of large models. Due to these challenges, it might be necessary to use on-demand VM instances for checkpointing on Spot VMs.

\textbf{Challenges in establishing an Economical VM Cluster including On-Demand VM:} 
The primary challenge in establishing an economical VM cluster, including on-demand options, lies in the complex and variable pricing structure of these resources. This complexity necessitates careful evaluation to find the best cluster configuration that not only fits within budgetary constraints but also maximizes efficiency in various dimensions~\cite{clustering-analysis2023}. 
This is crucial because, while GPU-based on-demand VMs are typically more expensive and thus pose a financial challenge, CPU-based on-demand VMs can sometimes be comparably priced or even cheaper than Spot GPU-VMs. Therefore, users must consider not only the technical differences between GPU and CPU VMs but also the economic implications of each choice. 

\textbf{Limitations of Existing Approaches:}
There is a very small 
effort 
to construct cost-effective clusters that leverage both Spot and On-Demand VMs. Some studies~\cite{Proteus,spotnik,DeepSpotCloud,andrzejak2010decision} are limited by their narrow applicability, as they are effective only in certain usecases and/or are tailored for specific architectures. Other approaches~\cite{wang2010distributed,deochake2023cloud,kokkinos2013cost,malta2019exploring} 
fail adequately propose methods for cluster configuration, lack comprehensive analysis or experimentation in complex scenarios, and fail to simultaneously consider both price and performance.

Therefore, this study introduces \tbdfixed{}, a novel approach toward cost-effective cluster configurations by intelligently balancing the use of Spot and On-Demand VMs. This operates through a four-stage process, (i) \tbdfixed{} begins by collecting the user pricing willingness; 
(ii) analyzing all available instances, using the ‘FLOPP' (Floating-point Operations Per Price) metric to assess each instance's performance in terms of its cost; (iii) the performance of each instance is evaluated within predefined architectures. This stage employs linear programming to examine the most cost-effective combinations within each architecture, taking into account constraints and potential overheads; (iv) 
identifies the best combination, culminating the process by choosing the most efficient and economical configuration for user-specific needs.
\tbdfixed{} achieves cost-efficiency, performance consideration, and flexibility through these stages.

We evaluated \tbdfixed{} using simulations and a real deployment in the AWS environment.
For the simulation, we experimented with seventeen different types of virtual instances. 
The results show that the configurations recommended by \tbdfixed{} consistently outperformed the other policies across all price ranges. For real-world 
AWS environment, we performed experiments on ten available instance types available at that time. The configurations recommended by \tbdfixed{} exhibited lower total costs and shorter makespan compared to other policies.
\section{Background and Related Work}
\label{sec:background}

In this section, we will discuss the necessity of Cloud Spot VMs, the importance of checkpointing, and the limitations of existing relevant research.

\subsection{Distributed Deep Learning}

In general deep neural network (DNN) training, the use of a single device such as a GPU is typically used to accelerate the training process.
However, when dealing with large-scale datasets or larger DNN models, a single computational device is insufficient.
To address this lack of computation resources, DDL techniques are employed, i.e., utilizing multiple GPUs and machines to divide and conquer the workload.
This approach enables parallel processing while the improving training efficiency and allowing for large-scale training.

In particular, when the size of a single data or mini-batch size is large, it may lead to out-of-memory errors on a GPU device, which makes it impossible to process a \textit{step} for a single mini-batch.
In the case, a mini-batch is divided and distributed across multiple devices.
This technique of partitioning a mini-batch into multiple GPUs and processing them in parallel is known as data parallelism.

In this context, 
a critical aspect is the synchronization of gradients across multiple GPUs. This synchronization is typically achieved using allreduce method. Recently, more efficient variations of allreduce have been developed to mitigate its impact on training speed. These include tree-based~\cite{Massively_Scale_DL} and ring-based~\cite{NCCL} allreduce structures, which offer more efficient communication patterns than the point-to-point (P2P) methods commonly used in parameter server approaches~\cite{MU_Li_scale_PS}. In this work, we particularly focus on a ring-based allreduce structure for a DDL workload.

\subsection{Preemption Hazard of Cloud Spot VM}

CSP offers computing resources to users by providing various VM instance types.
To maximize profits by utilizing idle resources more effectively, CSPs provide unclaimed or low-demand computing resources in the form of Spot VM instances.
Users purchase this type of computing resource in the Spot market which is typically managed through a bidding system.
Prices are dynamically adjusted based on supply and demand; therefore they 
fluctuate more frequently.
Spot market offers users cost-effective options because of its low pricing or even free usage compared with a normal on-demand VM of the same specification~\cite{CAI201838}.

Spot VMs are subject to cost fluctuations in cloud service prices, which poses certain challenges.
Depending on the policies of the CSPs, Spot VM instances can be preempted at any time when there is high demand or supply pressure for these resources.
There is an advantage to users as they can use these resources at a very low cost; however, users need to be careful of the fact that Spot VM instances are temporarily allocated to them and can be reclaimed by a CSP unexpectedly.
For example, in the case of Google Cloud Platform (GCP), preemption is reported to occur at least once every 24 hours~\cite{narayanan2020analysis}.

Preemption refers to the termination of a Spot VM instance by CSP when the demand for resources exceeds the available supply.
When a Spot VM instance is preempted, any data stored in the local memory and local disk are lost.
This is analogous to a hardware failure experienced by a typical server.

\subsection{Spot VMs and the On-Demand Dilemma in Checkpointing}
In distributed deep learning (DDL), users utilize checkpoints to periodically save training parameters such as DDL models and optimizer states.
These checkpoints serve multiple purposes.
Firstly, they assist in resuming training from the point of interruption due to system failures or unexpected terminations~\cite{eval-multi-level,CheckFreq,DeepFreeze,Wood-PyMM-ckp,check-n-run, Gemini, opt-ckp-restart,Desh}.
Secondly, they contribute to comparing the performance of models at specific epochs, aiding in the selection of the optimal model~\cite{Boost_by_CKP,ckp-best,CKP-Ensembles}.

\begin{table}[hbt!] 
\footnotesize
    \vspace{-10pt}
    \centering
    \caption{\small In-house testbed storage node specifications.}
   \vspace{-6pt}
    \label{tbl:machine}
    \resizebox{0.48\textwidth}{!}{
       \begin{tabular}{|l||l|}
            \hline
            Processor & AMD Ryzen 9 3900XT 12-Core Processor \\ \hline
            Main Memory  & DDR4, 16 GB x 4 (= 64 GB) \\ \hline

            OS Kernel & Linux version 5.15.0-72-generic \#79-Ubuntu SMP \\ \hline
            Network File System & ext4,  mount point: /mnt/nfs - 464.7 GB \\ \hline
            Storage Device & Samsung SSD 970 EVO 500GB \\ \hline
            
        \end{tabular}
    }
\end{table}

Thirdly, they are beneficial in addressing various issues such as debugging, and regression analysis that may arise during the deep learning process~\cite{ckp_for_DL,Error-CKP,CKP_Analysis}.
All these purposes significantly contribute to enhancing the efficiency and effectiveness of DDL training.

\begin{figure}[!t]
\centering
    \begin{tabular}{@{}c@{}c@{}c}
        \includegraphics[width=0.2\textwidth]{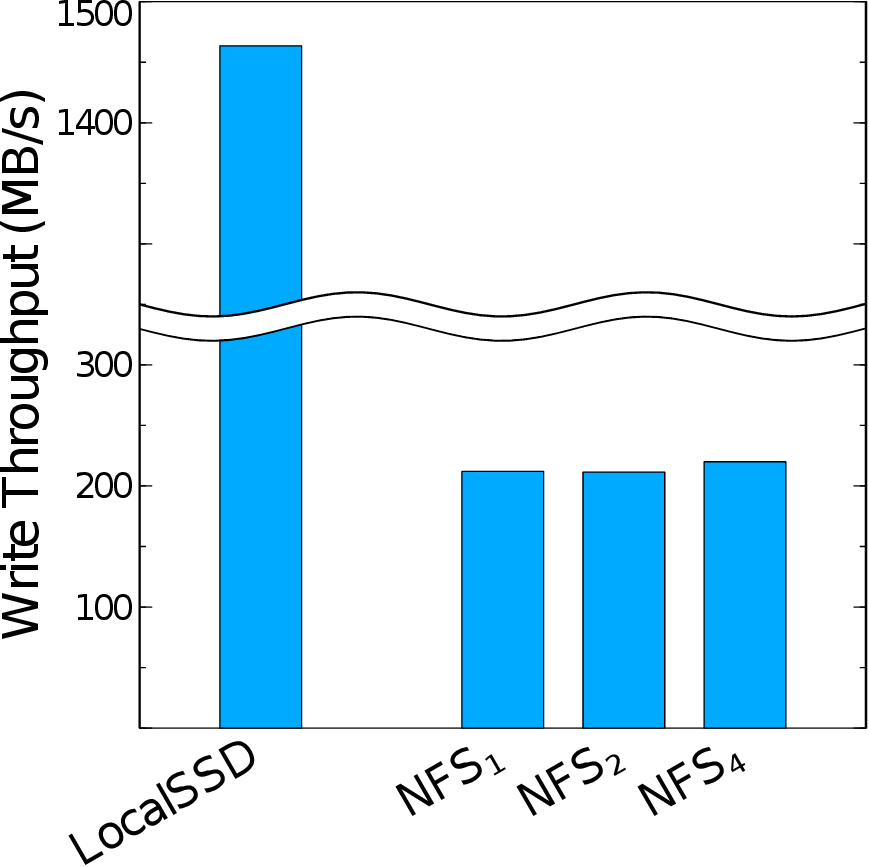} &
        \hspace{5pt}
        \includegraphics[width=0.2\textwidth]{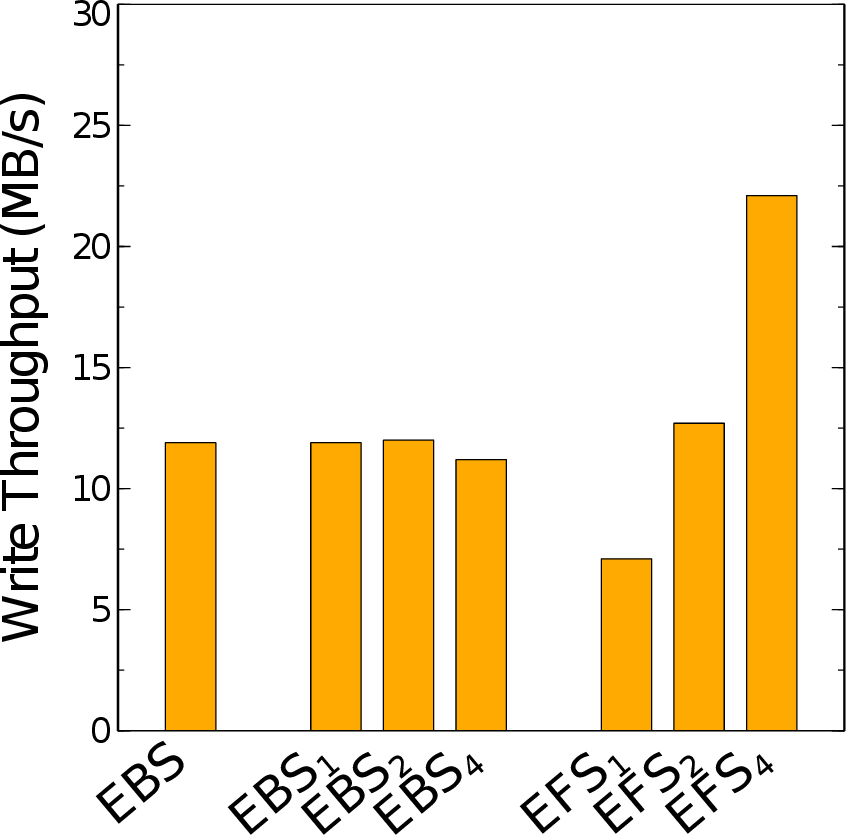} \\
        \scriptsize (a) Testbed Storage&
        \scriptsize (b) Cloud Storage\\
    \end{tabular}
\caption{Write throughput measurements of in-house testbed and AWS storage. We used 16 writer threads. with each thread performing 4KB block writing. The execution time for each experiment was set to 60 seconds.
}
\vspace{-10pt}
\label{fig:motivation}
\end{figure}
\index{figure}

Nevertheless, using checkpoints in Spot VM clusters presents several challenges.
The primary issue is the inaccessibility of Spot VM instance storage upon preemption, rendering local storage unusable for checkpoint storage. 
Consequently, users must consider alternative storage solutions.
One immediate alternative is the use of external storage; however, this too comes with its own set of challenges.
For example, when using external storage, write speed issues can occur.

{To empirically support the claim that cloud I/O performance is magnitude lower compared to on-premises local testbed, we performed write performance evaluation using FIO~\cite{fio} benchmark across various storage configurations. For our local in-house testbed, we used two storage setups: a local SSD device with an ext4 file system mounted, and an SSD mounted with NFS utilizing the NFSv4 protocol. The hardware specifications are listed in Table~\ref{tbl:machine}. In the case of the NFS configuration, we measured the aggregate write throughput by simultaneously writing to the storage from a single node, two nodes, and four nodes. Similarly, for evaluating the write performance of AWS storage, specifically Elastic Block Store~(EBS) and Elastic File System~(EFS), we used a g4dn.xlarge GPU VM instance on AWS.}



According to Figure~\ref{fig:motivation}, the write speed of external EBS is considerably slower than local SSDs, which can slow down or even halt the learning process due to I/O. Since the preemption time of Spot VMs is fundamentally unpredictable, rapid data storage becomes crucial.
Unfortunately, the slow write speed of EBS fails to meet this requirement. Alternatively, using external EFS incurs costs based on storage capacity and data transfer volume, making it an expensive solution, particularly when large models are checkpointed at each epoch.
This is not an economically viable option for Spot VM cluster users seeking cost-effective solutions.

To address these problems, users can opt for on-demand instances. On-demand instances are not subject to preemption, thereby ensuring the stability of checkpoint data and reducing the need for rapid checkpointing.
However, on-demand instances are more expensive than Spot VMs, and their use involves considering several variables.
Decisions such as whether to involve on-demand instances in training, use them solely as memory nodes, choose the type of instances, and determine the number of instances required, complicate the economic configuration of the cluster and present a significant challenge to users.

\subsection{Existing Approaches and Their Limitations}
\label{sec:limitation}

We observe several limitations in existing resource allocation research that make them unsuitable for constructing cost-effective clusters leveraging both Spot and On-Demand VMs.

\noindent\textbf{Narrow Applicability:}
Some existing approaches often suffer from limited applicability, restricting their effectiveness in specific scenarios. 
Proteus~\cite{Proteus} 
only considers distributed learning environments with a Parameter Server (PS) architecture, and as the number of Spot VMs increases, the proportion of On-Demand VM usage also increases, reducing cost-efficiency.
Spotnik~\cite{spotnik} 
is limited to situations where only part of the cluster faces preemption and is incapable of avoiding complete checkpoint approaches when the entire cluster is revoked.
DeepSpotCloud~\cite{DeepSpotCloud} 
is focused only on single GPU instances and lacks the necessary expansion for distributed deep learning platforms.
Andrzejak et al.~\cite{andrzejak2010decision} 
only consider Spot VMs and does not consider On-Demand VMs.

\noindent\textbf{Inadequate Methods for Cluster Configuration:}
Some existing approaches fail to adequately propose methods for cluster configuration or have limitations in their proposed approaches.
H. Wang et al.~\cite{wang2010distributed} analyze the relationship between cloud pricing and system configurations, but lacks a method for how best to configure clusters based on this analysis and does not experiment with it.
S. Deochake~\cite{deochake2023cloud} investigates various methods for optimizing cloud costs, acknowledging the potential for cost savings. However, it lacks analysis in complex scenarios, particularly in configuring clusters considering both price and performance simultaneously.
P. Kokkinos et al.~\cite{kokkinos2013cost} perform resizing based on the utilization and cost of currently used instances. However, it fails to propose a cluster configuration prior to performing tasks, limited to configuring clusters only through monitoring.
E. M. Malta et al.~\cite{malta2019exploring} identify cost-effective instances for deep learning applications based on the cost and performance of instances and presents a general methodology. However, this approach requires predicting performance and cost for each application, leading to the disadvantage that the most efficient instance may vary for each application.

\section{Design of \tbdfixed{}}

In this section, we first introduce the necessity of an economical cluster configuration for distributed deep learning. Following that, we provide a detailed explanation of the design and implementation of DeepVM proposed in this study.

\subsection{Challenges}

Our primary goal is to propose an effective cluster architecture and configuration for using Spot VMs and checkpoints in cloud environments. Some users believe that constructing a cluster with the cheapest Spot VMs is economical. Others think that using the highest-spec instances as Spot VMs for faster training will yield economic benefits. Still, others seek a balance between price and performance. 
A common goal among all these users is to efficiently conduct deep learning using Spot VMs in a cost-effective cluster. 
However, forming a cluster based solely on current prices is challenging. 
This challenge is compounded when considering On-Demand instances for checkpointing strategies.

The key insight of our research is the vast array of variables to consider in cloud-based distributed deep learning and the significant price differences among VM types. 
Table~\ref{tbl:awstypes} shows various VM specifications and prices, illustrating the difficulty for users to determine which VM is cost-effective. This poses a major consideration in complex cluster configuration decisions. However, the variety of instances also presents a challenging dilemma in forming an economical cluster. 
Therefore, to achieve the optimal cluster configuration, the following challenges need to be addressed.

\begin{table}[!t]
\small
\scriptsize 
\caption{\small Various Types of instances provided by AWS.
\\This information was available on October 1, 2023 in N.Virginia region.
}
\vspace{-6pt}
\label{tbl:awstypes}
\resizebox{0.48\textwidth}{!}{
    \begin{tabular}{@{}c@{}|c|c|c|c|c|c|c@{}}
    \toprule[1.5pt]
    \multirow{2}{*}{Name} & \multirow{2}{*}{Type} & On- & Spot & vCPU & Mem & Network & Storage           \\ 
            &  &  Demand &  VM & (\#) & (GiB) & (Gbps)    & (Gbps)       \\ \hline \hline
    g3s.lxarge & GPU & \cellcolor{yellow!27}0.75 & \cellcolor{green!15}0.225 & 4 & 30.5 & - 10 & N/A \\ \hline
    g4dn.xlarge & GPU  & \cellcolor{yellow!20}0.526 &\cellcolor{green!14.5}0.1941 & 4 & 16 & - 25  & - 3.5    \\ \hline
    g5.xlarge & GPU & \cellcolor{yellow!35}1.006 &\cellcolor{green!20}0.3018 & 4 & 16 & - 10  & - 3.5  \\ \hline \hline
    p2.xlarge & GPU  &\cellcolor{yellow!13}0.2704 & \cellcolor{green!15}0.225 & 4 & 61 & High & 0.75 \\ \hline
    p2.8xlarge & GPU  &\cellcolor{yellow!62}7.2 & \cellcolor{green!55}2.1165 & 32 & 488 & 10  & - 5  \\ \hline
    p2.16xlarge & GPU & \cellcolor{yellow!100}14.4 & \cellcolor{green!100}4.2262 & 64 & 732 & 25  & - 10  \\ \hline \hline
    
    c3.xlarge & CPU & \cellcolor{yellow!11}0.21 &\cellcolor{green!5}0.1281 & 4 & 7.5 & Moderate & N/A \\ \hline    
    c4.xlarge & CPU & \cellcolor{yellow!11}0.199 & \cellcolor{green!5}0.1257 & 4 & 7.5 & High & 0.75 \\ \hline
    t3.xlarge & CPU & \cellcolor{yellow!10}0.1664 & \cellcolor{green!4}0.1033 & 4 & 16 & - 5  & N/A \\ \hline   
    c5.xlarge & CPU & \cellcolor{yellow!10}0.17 & \cellcolor{green!3}0.0999 & 4 & 8 & - 10  & - 4.75  \\ \hline
    m6i.xlarge & CPU & \cellcolor{yellow!10}0.192
    &\cellcolor{green!5}0.12 & 4 & 16 & - 12.5  & - 10 \\ \hline
    d3.xlarge & CPU & \cellcolor{yellow!20}0.499 &\cellcolor{green!5}0.1584 & 4 & 32 & - 15  & 0.85  \\ \hline
    c5n.xlarge & CPU & \cellcolor{yellow!12}0.216 &\cellcolor{green!5}0.127 & 4 & 10.5 & - 25  & - 4.75  \\ \hline
    c6in.xlarge & CPU & \cellcolor{yellow!12}0.2268 &\cellcolor{green!5}0.1272 & 4 & 8 & - 30  & - 20  \\ \bottomrule[1.5pt]
    
    \end{tabular}
}
\vspace{-15pt}
\end{table}

\squishlist
\item
First, determining the price-performance ratio for each instance can be challenging.
This involves comparing the performance and cost among various VM options to identify the optimal configuration.
\item
Second, considering overheads in parallel processing, it is essential to focus on two primary factors. Firstly, the challenge lies in scaling distributed learning efficiently. This requires a careful selection of instance types and numbers to ensure that the system can grow without losing effectiveness. Secondly, preventing network saturation is crucial. An optimal balance must be maintained in instance variety and quantity to avoid overloading the network.
\item
Third,  
determining whether On-Demand instances should be actively involved in training or used primarily as memory nodes for checkpointing processes is the third challenge. This is a critical aspect in achieving the most efficient and cost-effective use of resources in distributed deep learning.
\squishend

\label{sec:overview}
\subsection{Overview}

\tbdfixed{} operates through the following four stages: User Pricing Input, Instance-level Analysis, Architecture-level Analysis, and Final Decision.

\begin{enumerate}
\item \textbf{Step\#1:} User Pricing Input: \tbdfixed{} begins with collecting the user's Pricing Willingness. In this initial phase, the algorithm gathers data on the maximum price per hour that the user is willing to pay.

\item \textbf{Step\#2:} Instance-level Analysis: This stage analyze all available instances for the user. Using the `FLOPP' metric, it measures each instance's performance relative to its cost.

\item \textbf{Step\#3:} Architecture-level Analysis: Upon completing instance analysis, the performance of each instance in predefined architectures is analyzed. This stage explores the most cost-effective instance combinations within these architectures using linear programming, considering each architecture's constraints and potential overheads in parallel processing. 
The paper selects two architectures as representative examples: 
`Single Anchor', a combination of one on-demand GPU-VM and multiple spot GPU-VMs, and `Tiering', a combination of on-demand CPU-VMs as memory nodes and spot GPU-VMs for training.


\item \textbf{Step\#4:} Final Decision: After identifying the best combinations for all architectures, this stage compares their performance to propose the most cost-effective cluster configuration under the given conditions.
\end{enumerate}

\begin{figure}[!t]
    \centering  \includegraphics[width=0.48\textwidth]{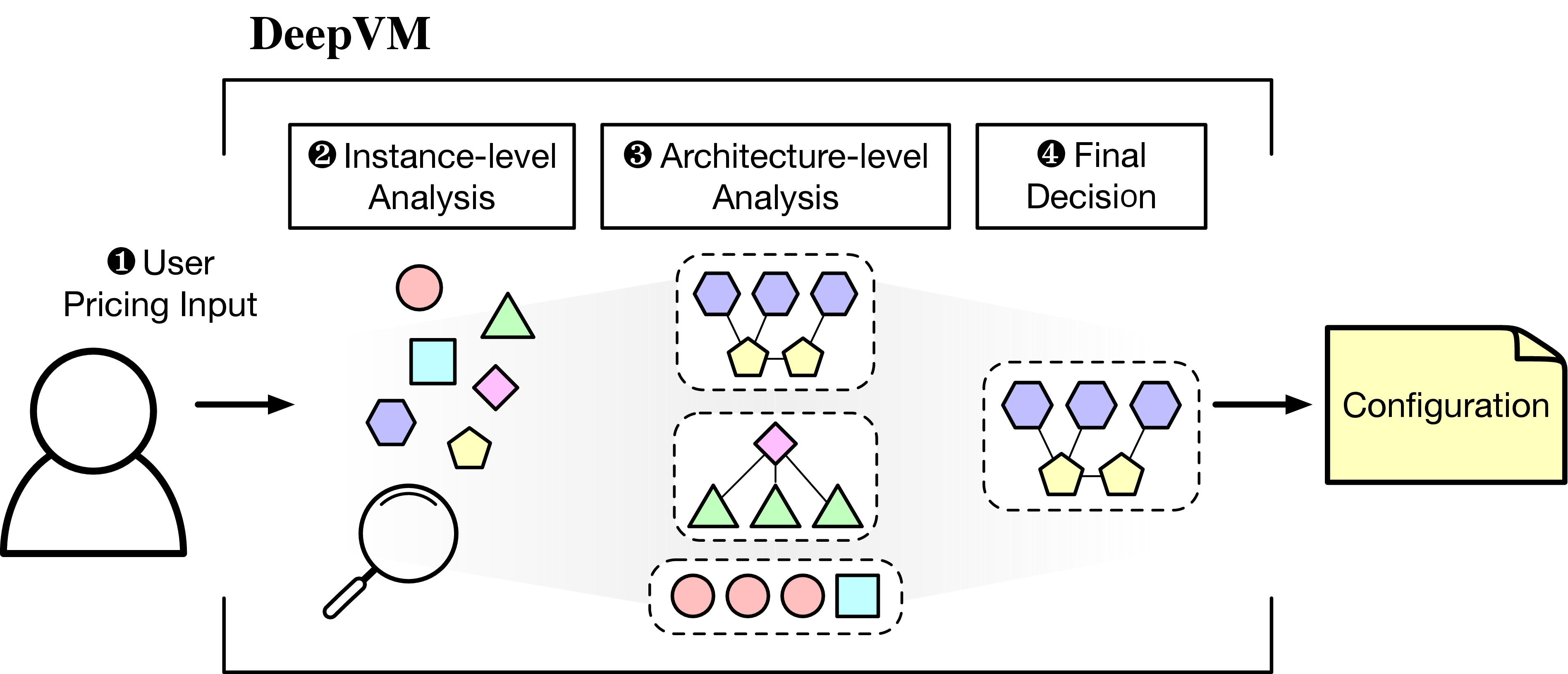}
   \vspace{-5pt}
    \caption{An overview of \tbdfixed{}.}
    \vspace{-13pt}
    \label{fig:Overview}
\end{figure}

Figure~\ref{fig:Overview} describes these steps of \tbdfixed. Table~\ref{table:notation} defines the mathematical notations used in the algorithm.

\subsubsection{User Pricing Input}
In the initial phase of the \tbdfixed{} algorithm, the user is prompted to specify their Pricing Willingness (PW). This figure represents the maximum amount, on an hourly basis, that the user is willing to pay. The user defines their financial threshold for the cloud resources by setting their maximum affordable hourly rate. This rate is pivotal in guiding the subsequent stages of the algorithm. Based on the primary intention of forming an economical cluster using spot instances, it is assumed that users will typically propose a price lower than what a fully on-demand cluster would cost. This assumption aligns with the goal of achieving cost-efficiency in a cloud environment. The flexibility in PW allows users to tailor their experience. Those seeking higher performance might opt for a higher PW, while users prioritizing cost savings might set a lower PW. This variability ensures that the algorithm caters to a diverse range of user needs and preferences.  Throughout the subsequent stages, the algorithm utilizes the PW as a key parameter. It strives to maximize performance relative to cost, aiming to recommend the most cost-effective cluster configuration that aligns with the user's set PW.

\begin{table}[t!]
\begin{center}
\scriptsize
\caption{Declaration of Variables}
\vspace{-6pt}
 \begin{tabular}{|l|p{6.8cm}|} 
 \hline
 Variable & Description \\ [0.5ex] 
 \hline\hline
 \multicolumn{2}{| c |}{Input}\\
 \hline
    $\mathbf{V}$  & Set of GPU-VM instances \\ 
    $v$  & GPU-VM instance \\
    $\mathbf{W}$  & Set of CPU-VM instances \\
    $w$  & CPU-VM instance \\
    $PW$ & Pricing willingness \\
    $f$  & Size of the checkpoint file \\
    $s$  & Buffer size of remote memory \\
    $\mathbf{A}$  & Set of VM cluster architectures and their constraints and performance functions. \\ [0.5ex]
    \hline \hline
    \multicolumn{2}{| c |}{Output}\\
    \hline
    $a_0$ & Recommended architecture \\ 
    $v_0$ & Recommended GPU-VM instances \\
    $n_0$ & Number of $v_0$ \\
    $w_0$ & Recommended CPU-VM instances \\
    $m_0$ & Number of $w_0$ \\
    $P_0$ & Hourly price \\ [0.5ex]
    \hline \hline
    \multicolumn{2}{| c |}{Element of Instance Tuple}\\
    \hline
    $SPFP(x)$ & Spot VM $FLOPP$ of instance $x$ \\
    $ODFP(x)$ & On-demand VM $FLOPP$ of instance $x$ \\
    $SPPR(x)$ & Spot VM price per hour of instance $x$ \\
    $ODPR(x)$ & On-demand VM price per hour of instance $x$ \\
    $MEM(x)$ & Memory capacity of instance $x$ \\
 \hline
 \end{tabular}
\label{table:notation}
\vspace{-20pt}
\end{center}
\end{table}

\subsubsection{Instance-level Analysis}
To analyze the cost-efficiency of individual instances, understanding how to simultaneously consider price and performance is essential. For performance, one might consider the theoretical FLOPS (Floating Point Operations Per Second) of GPUs. However, actual deep learning performance can differ due to factors like data transfer and memory access patterns. 
To address this, we introduced `eFLOPS', a numerical metric representing performance in executing deep learning workloads. We specifically selected image processing workloads as representative tasks for deep learning, benchmarking them across various AWS instances to reflect real-world performance.
{Additionally, users or administrators interested in different tasks can benchmark performance using execution time.}
Parallel to this, we analyzed the hourly pricing of instances provided by Cloud Service Providers (CSPs) to consider the cost aspect. This led to the derivation of `FLOPP' (Floating-point Operations Per Price), {a metric defined in Equation \ref{eq:flopp}.}
\begin{align}
\label{eq:flopp}
\textrm{FLOPP} := \textrm{eFLOPS} \div \textrm{{(price per time)}} \nonumber \\\
= \frac{operations}{time} \times \frac{time}{price} = \frac{operations}{price}
\end{align}
FLOPP indicates the amount of deep learning operations that can be performed per unit price, which varies based on spot and on-demand options, and price fluctuations.
\tbdfixed{} pre-calculates the eFLOPS value for each instance. Utilizing this, the algorithm computes the on-demand and spot FLOPP values for available instances to determine the most optimal instance selection.

\subsubsection{Architecture-level Analysis}
\label{previous-section}
Following the instance evaluation, \tbdfixed{} considers a wide and expandable range of predefined architectures
$\mathbf{A}$ (e.g., Single-Anchor or Tiering Architecture) to find the optimal combination of instances. For simplicity, we assume all GPU-VMs use the same instance. The performance of an architecture is represented by the product of the performance of the GPU-VMs involved in training and their quantity. We account for different FLOPP values between spot and on-demand instances and introduce a $ScalingFactor$ to reflect the non-linear performance increase in parallel tasks. Thus, the performance function Z, as described in Equation \ref{eq:perf_function_z}, models this relationship.
\begin{equation}
\label{eq:perf_function_z}
Z=\left(N \times SPFP(v) + M \times ODFP(v) \right) \times ScalingFactor(v,n) 
\end{equation}
{where $N$ is the number of spot instances, and $M$ is the number of on-demand instances.}
The algorithm determines the combination of instances that maximize this function using linear programming. Detailed explanation of $ScalingFactor${, as defined in Equation \ref{eq:scaling_factor},} can be found in Section~\ref{sec:scaling_modeling}.

\begin{figure}[!t]
    \centering  \includegraphics[width=0.48\textwidth]{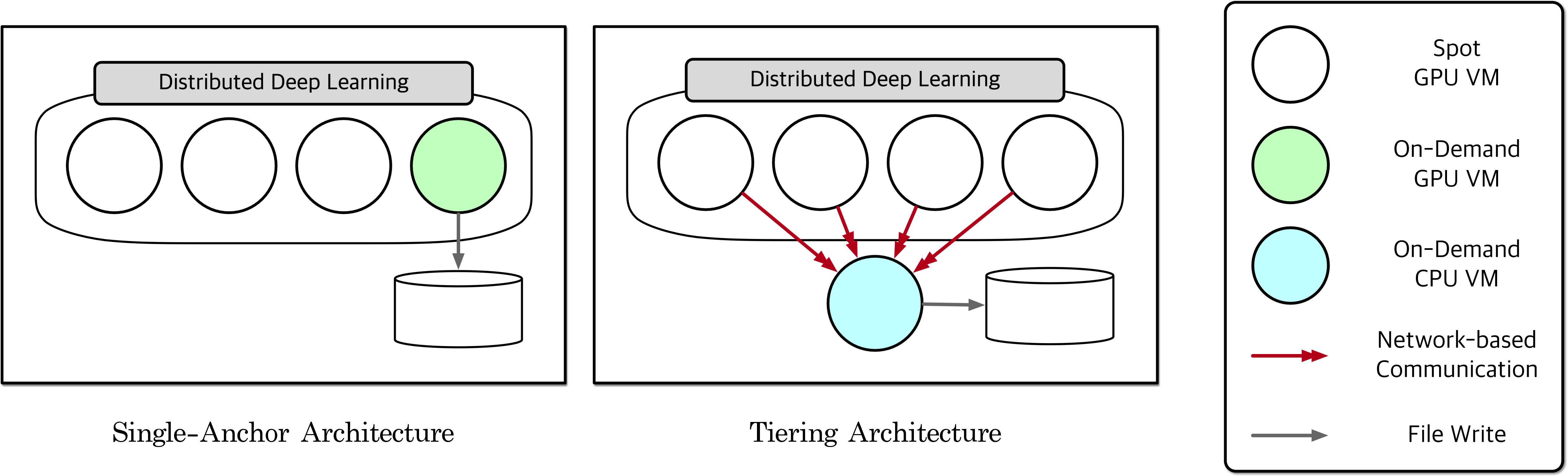}
    \vspace{-6pt}
    \caption{Single Anchor and Tiering Architecture.}
    \vspace{-15pt}
    \label{fig:Architecture}
\end{figure}

Figure~\ref{fig:Architecture} illustrates the two architectural scenarios assumed in this paper.
{Two representatively selected predefined architectures}
are `Single Anchor' and `Tiering'. The description and constraints of each are as follows:
\squishlist
\item In the `Single Anchor' architecture, one on-demand and the remaining spot instances are used among $n$ total participating instances. {Here, $N$ represents $n-1$ (the number of spot instances) and $M$ is 1 (the single on-demand instance) in the context of Equation \ref{eq:perf_function_z}.} The on-demand VM is used for checkpointing, and the total hourly cost of the cluster must not exceed the user's pricing willingness ($PW$).
{Therefore, the constraints, as formalized in Equation~\ref{eq:cons_single}, dictate the following relationship.}
\begin{equation}
\label{eq:cons_single}
(n-1)\times SPPR(v)+ ODPR(v) \le PW
\end{equation}
Under these constraints, {the maximization of the performance function Z is represented in Equation \ref{eq:z_single}.}
\begin{equation}
\label{eq:z_single}
Z=((n-1)\times SPFP(v)+ODFP(v))\times ScalingFactor(v,n)    
\end{equation}
\item In the `Tiering' architecture uses $n$ spot GPU-VM instances for training and $m$ on-demand CPU-VMs for multi-level checkpointing. This architecture focuses on the cost-efficiency of on-demand CPU-VMs. Users can use this structure with software supporting multi-level checkpointing. {In this setup, $N$ equals $n$ and $M$ equals $m$ in the context of Equation \ref{eq:perf_function_z}.} We assume that users are utilizing software that supports multi-level checkpointing and parallel transmission of model parameters. Adequate memory in CPU-VMs and proper ratio adjustment between GPU-VMs and CPU-VMs are crucial to prevent network bottlenecks, keeping the hourly cost below $PW$. 
{Hence, the constraints, outlined in Equations~\ref{eq:cons_tiering1},~\ref{eq:cons_tiering2} and~\ref{eq:cons_tiering3}, are as previously described.}
\begin{align}
\label{eq:cons_tiering1}
    n \times SPPR(v) + m \times ODPR(w) \le PW \\
\label{eq:cons_tiering2}
    MEM(w) \ge f \times s \\
\label{eq:cons_tiering3}
    \frac{n}{m} < NWSaturationPoint(v,w)
\end{align}
Under these constraints, the performance function Z is maximized {as depicted in Equation \ref{eq:z:tiering}.}
\begin{equation}
\label{eq:z:tiering}
Z=n\times SPFP(v)\times ScalingFactor(v,n)
\end{equation}
Details on $NWSaturationPoint${, as well as its implications in the context of Equations~\ref{eq:tiering_policy},} are discussed in Section~\ref{sec:network_modeling}.
\squishend

{If additional architectures are added to the set of predefined architectures $\mathbf{A}$, constraints and performance functions specific to these architectures will also be included, allowing \tbdfixed{} to consider a broader range of architectures.}


\subsubsection{Final Decision}
Upon completing the analysis of architectures, the maximum values of the performance function Z for each architecture are compared. This process selects the architecture that provides the best performance within the given pricing willingness, ensuring the highest cost-efficiency. The selected architecture is then presented to the user, who has the option to utilize it or return to the initial step with a modified PW value, considering different price or performance requirements.

\subsection{Overhead Modeling}
\label{sec:overhead}

\subsubsection{Scaling factor of multiple GPUs}
\label{sec:scaling_modeling}

Distributed deep learning applications do not exhibit a linear increase in the number of operations processed per unit time by GPUs due to overheads such as data loading and synchronization. 
When deploying distributed deep learning applications using multiple GPU-VMs, it is crucial to consider the mentioned overheads to accurately predict performance based on the number of GPU-VMs.

\tbdfixed{} does not simply represent the cluster's overall performance as a naive linear extension with respect to the number of instances ($n$). 
Instead, it constructs a prediction function using a measurement-based linear regression technique that accounts for overheads, enabling performance prediction that includes these factors.

For \tbdfixed{}, we measured the execution time of distributed deep learning workloads by progressively increasing the number of GPU-VMs comprising the VM cluster and calculated the speedup.
We assume that each VM instance in the VM cluster is equipped with one GPU.
Speedup quantifies how much the computational speed increases when we scale the number of GPU-VMs by a factor of $n$.
The value of $K$ considered within the cluster's overall performance is ultimately closely related to this speedup. 
{$K$ is expressed in Equation~\ref{eq:K}.}
\begin{equation}
\label{eq:K}
K(n)=\frac{(\textrm{actual speedup})}{(\textrm{theoretical speedup})}
\end{equation}

Through experiments, we observed that the performance of a specific DDL workload exhibits linear speedup when the number of GPU-VMs is low, but transitions into a logistic function shape. 
This transition aligns with our expectations, as communication complexity increases in the distributed environment, leading to amplified overhead.
To model this, we used logistic regression as formulated {in Equation~\ref{eq:regression}}.

\begin{equation}
\label{eq:regression}
S(workload, n) = \frac{c}{1+\exp(-a(n-b))}
\end{equation}

\begin{figure}[!t]
    \centering  \includegraphics[width=0.48\textwidth]{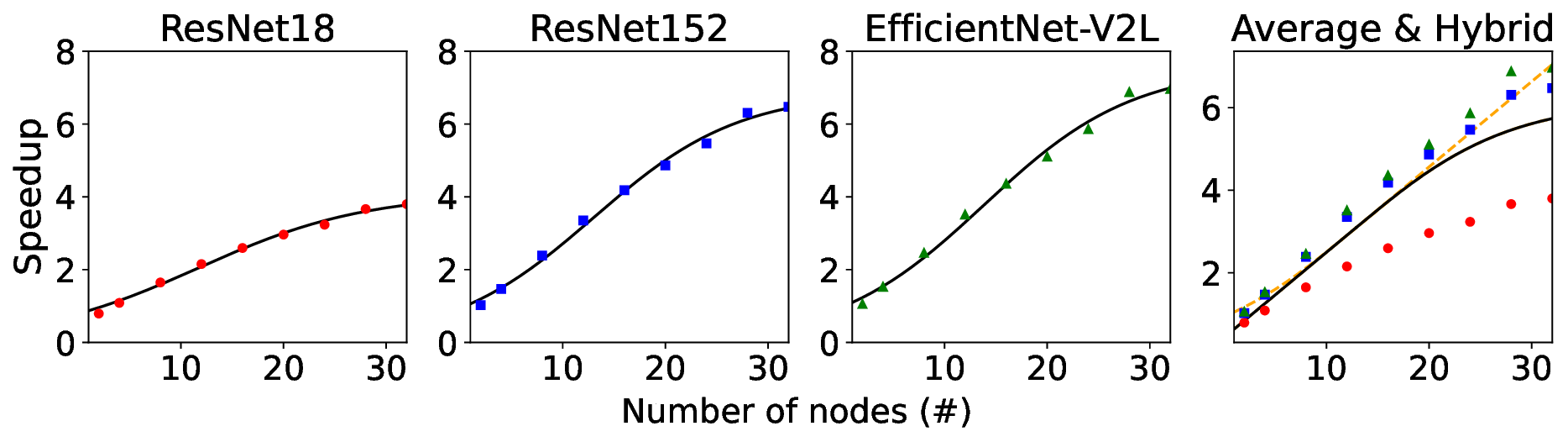}
    \vspace{-10pt}
    \caption{
    Speedup results as the number of GPU-VMs increased for three different DL image models. g4dn.xlarge VMs were used.
    }
    \label{fig:Speedups}
    \vspace{-8pt}
\end{figure}

\begin{table}[!t]
\centering
\scriptsize
\caption{
Regression results for speedup of each model.}
\resizebox{0.34\textwidth}{!}{
\begin{tabular}{|l|c|c|c|}
\hline
Model Name & $a$ & $b$ & $c$ \\
\hline
\hline
ResNet18 & 0.1222 & 11.7094 & 4.0927 \\
ResNet152 & 0.1414 & 13.0476 & 6.8803 \\
EfficientNet-V2L & 0.1380 & 13.8657 & 7.5652 \\
\hline
Average & 0.1339 & 12.8742 & 6.1766 \\
\hline
\end{tabular}
}
\vspace{-15pt}
\label{tab:regression}
\end{table}

We selected prominent image processing models, namely ResNet18, ResNet152, and Efficient-Net, and performed these workloads on GPU-VM instances.
Figure \ref{fig:Speedups} and Table \ref{tab:regression} present the logistic regression results when executing these three workloads on AWS's $\mathtt{g4dn.xlarge}$.
From these results, we derived the values $a$, $b$, and $c$ for each model. 
We then obtained $\bar{a}$, $\bar{b}$, and $\bar{c}$ by averaging these values.
We introduced a robust average speedup metric, denoted as $S_{average}${, which is defined in Equation~\ref{eq:average_S}}.
\begin{equation}
\label{eq:average_S}
S_{average}(n)=\frac{\bar{c}}{1+exp(-\bar{a}(n-\bar{b}))}
\end{equation}

$S_{average}$ closely follows the empirical speedup trend for sufficiently large values of $n$; however, it may not accurately represent linearity when $n$ is small. 
Additionally, these trend functions can vary for different workloads.
In our research, as previously mentioned, we only considered deep learning workloads based on image processing models.

On the one hand, this function exhibits an inflection point at $n=\bar{b}$. 
Let's denote the tangent line at this inflection point as $L(n)$.
Since $L(n)$ serves as a good linear approximation of the original function, it effectively captures the linear characteristics of the initial speedup. 
Therefore, to achieve a modeling that better satisfies experimental trends, we define $S_{hybrid}$ using $L(n)$.
{This is expressed in Equation~\ref{eq:s_hybird}}.

\begin{equation}
\label{eq:s_hybird}
S_{hybrid}(n) = 
\begin{cases}
L(n) & n\leq \bar{b}
\\ S_{average}(n) & n>\bar{b}
\end{cases}    
\end{equation}

On the theoretical side, it can be considered that the speedup of $n$ instances follows $y(n) = n$.
Ultimately, {$K$ is defined using $y(n) = n$ and $S_{hybrid}$, as shown in Equation~\ref{eq:scaling_factor}}.

\begin{equation}
\label{eq:scaling_factor}
K(n)=\frac{S_{hybrid}(n)}{n}
\end{equation}

The value of $K$ is considered as a scaling factor to account for overhead in the cluster's overall performance.
Therefore, the relative performance when using $n$ GPU-VM instances ($X$) is obtained by multiplying $X$'s FLOPP by $n$ and $K(n)$. 
This allows for performance measurements that take into account the complexity and overhead when using multiple GPU-VMs.
\tbdfixed{} has a function table that contains information as described above. This table defines $S_{hybrid}$ for AWS's prominent GPU-VM instances.
\tbdfixed{} can reference this table to provide users with the optimal configuration.

\subsubsection{Saturation Point of Network Bandwidth}
\label{sec:network_modeling}

We previously assumed a method of splitting model parameters into multiple shards (pieces) for transmission during multi-level checkpointing
(refer to Section~\ref{previous-section}). 
This approach facilitates faster checkpointing. However, when splitting a single checkpoint data into too many shards, it can lead to even slower transmission speed due to network saturation.

Let $t_{XY}(n)$ represent the time which is taken for $n$ GPU-VM instances ($X$) to transmit checkpoint data to a single CPU-VM instance ($Y$). 
Although users may expect the benefit of sharding to result in $t_{XY}(n) > t_{XY}(n+1)$, in reality, there exists an $n$ overhead where this expectation may not hold due to communication.

\begin{table}[!t]
\begin{center}
\scriptsize 
\caption{
values of $n_{sat}$ at which network bandwidth begins to saturate when multiple instances $X$ communicate N to 1 with one instance $Y$.
}
\resizebox{0.48\textwidth}{!}{
    \begin{tabular}{|l|c|c|c|c|c|c|c|c|}
        \hline
        $bandwidth$ & 0.3 & 1.7 & 5 & 10 & 12.5 & 15 & 25 & 30 \\
        \hline
        $n_{sat}$ & 3 & 12 & 16 & 20 & 24 & 24 & 28 & 32 \\
        \hline
    \end{tabular}
}
\label{tab:n_sat}
\vspace{-18pt}
\end{center}
\end{table}

To avoid from taking high communication overhead, if increasing the number of sharding nodes actually leads to a disadvantage in terms of time cost, \tbdfixed{} considers the communication overhead to be too high.
It meaning $t_{XY}(n)$ is less than or equal to $t_{XY}(n+1)$, 
{The definition of this policy is given by Equation~\ref{eq:network_policy}}.

\begin{equation}
\label{eq:network_policy}
t_{XY}(n) \le t_{XY}(n+1)
\end{equation}

Now, let's denote the minimum value of $n$ for which the above inequality holds, i.e., the value of $n$ at which disadvantage begins.
We'll set this value $n_{sat}(X)$.

For the minimum value of $n$ that satisfies the above inequality, i.e., when $n$ instances of $X$ simultaneously communicate with one instance of $Y$, and overhead starts occurring, we'll also denote it as $n_{sat}(X)$.

{In this context, the tiering policy, as defined in Equation~\ref{eq:tiering_policy}, posits that the ratio of GPU-VMs to CPU-VMs should be less than $n_{sat}:1$.}
\begin{equation}
\label{eq:tiering_policy}
    \frac{GPU_{VM}}{CPU_{VM}} <n_{sat}
\end{equation}

We sampled some instances from AWS to determine the bandwidth and the corresponding $n_{sat}$, which we then applied into \tbdfixed{} as a function.
Table \ref{tab:n_sat} shows the applied values of $n_{sat}$.

{The accuracy of the two overhead models discussed in this section is addressed in Section~\ref{sec:overhead_expr}.}

\section{Implementation}

We developed \tbdfixed{} using Python (3.10.12). For the speedup functions required for provisioning, we conducted regression using the SciPy library (1.11.2) in Python. 
To evaluate \tbdfixed{}, we also implemented a simulation environment using the same version of Python. 
This ensures consistency in our development and evaluation processes, providing a reliable platform for testing and analyzing the performance of \tbdfixed{}. For validating the Tiering architecture in \tbdfixed{}, we developed a software supporting asynchronous multilevel checkpointing. This software was developed utilizing PyTorch (2.0.1+cu117) and the MPI library (OpenMPI-v5.0.0rc12). 
Additionally, to further validate \tbdfixed{}, we developed and incorporated other algorithms into the code. 
These algorithms are designed to rigorously test and demonstrate the capabilities of \tbdfixed{} under various conditions and scenarios. 
{Notably, the \tbdfixed{} showcases minimal complexity and remarkable speed in executing linear programming tasks. Once the variables are input, \tbdfixed{} rapidly makes decisions, reflecting its efficiency in handling computational tasks.}
The source code of \tbdfixed{}, along with the associated tools and algorithms, is publicly available for download at \cb{https://github.com/lass-lab/DeepVM}.
\section{Evaluation}
\label{sec:eval}

In this section, we first evaluate the overhead of our proposed simulation modeling. Second, we explore the effectiveness of \tbdfixed{} for various configurations with different prices of VM instances and the user's Pricing Willingness.

\subsection{Experimental setup}

\begin{table}[!t]
\small
\caption{\small Specifications of Simulated VMs, including 10 GPU-VMs and 7 CPU-VMs based on AWS specifications.
}
\vspace{-6pt}

\label{tbl:simulated_instances}
\scriptsize 
\renewcommand{\tabcolsep}{0.56mm}
\begin{tabular}{|@{}c@{}|c|c|c|c|c|@{}c|c|c|c|c|c|@{}}
\toprule[1.0pt]
\hline
\multirow{2}{*}{Name} & \multirow{2}{*}{Type} & On- & Spot & Network & \multirow{2}{*}{eFLOPS} & & \multirow{2}{*}{Name} & \multirow{2}{*}{Type} & On- & Spot & Network   \\ 
        & &  Demand &  VM & (Gbps) & && &  &Demand &  VM & (Gbps) \\ \hline \hline
 A & GPU & 0.75 & 0.225 & 10 & 100 &&K & CPU & 0.199 & 0.126 & 1.7 \\ \hline
 B & GPU & 0.526 & 0.158 & 25  & 377 &&L & CPU & 0.1664 & 0.103 & 5    \\ \hline
 C & GPU & 1.006 & 0.302 & 10  & 696 
 && M & CPU & 0.17 & 0.1 & 10    \\ \hline
 D & GPU & 0.55 & 0.165 & 15  & 700 
 &&N & CPU & 0.192 & 0.12 & 12.5   \\ \hline
 E & GPU & 0.368 & 0.11 & 1.7  & 100    
 &&O & CPU & 0.499 & 0.158 & 15
 \\ \hline
 F & GPU & 1.236 & 0.371 & 25  & 800 
 && P & CPU & 0.216 & 0.127 & 25
 \\ \hline
 G & GPU & 0.973 & 0.292 & 30  & 200
 && Q & CPU & 0.2268 & 0.127 & 30
 \\ \hline
 H & GPU & 0.252 & 0.076 & 5  & 150 
 &&- & & &  & 
 \\ \hline
 I & GPU & 1.622 & 0.487 & 30  & 900    
 &&- & & &  &
 \\ \hline
 J & GPU & 0.22 & 0.066 & 5  & 50    
 &&- & & & &
 \\  \hline
 
 \bottomrule[1.0pt]
\end{tabular}
\vspace{-10pt}
\end{table}

In order to comprehensively evaluate the \tbdfixed{} system, we implemented a two-pronged experimental approach that included both simulated and actual AWS environment setups. This methodology enabled us to not only test the system within the constraints and variability of real-world cloud infrastructure but also to extend our investigation into a controlled simulation that could encompass a wider array of instance types and configurations than those currently available on AWS. By employing this dual setup, we aimed to provide a robust validation of \tbdfixed{}'s performance across a spectrum of scenarios.

\noindent\textbf{Simulation Experiment Setup}:
Given the limited availability of specific instance types on AWS and restrictions on the number of instances, we faced constraints in our experimental setup. To overcome these limitations and validate a broader range of instances, we turned to simulations. Consequently, we created additional virtual instances, simulating 10 GPU-VMs and 7 CPU-VMs based on AWS specifications. Among these, instances D through J were specifically designed to meet our experimental objectives, offering a variety of performance, pricing, and parallel processing capabilities. This approach allowed us to explore more diverse configurations and their impact on performance. Detailed descriptions of these instances, including their individual characteristics, are documented in Table \ref{tbl:simulated_instances}.
In this setup, we use the following metric to assess \tbdfixed{}.
\squishlist
\item
\textbf{Performance:} This represents a simulation of the capability to process deep learning workloads. The value is calculated by multiplying the eFLOPS of the GPU-VMs and their quantity, along with the $ScalingFactor$ that indicates the efficiency of parallel processing. Through this, we can evaluate the performance of each configuration.
\squishend

\noindent\textbf{Real AWS Experiment Setup}:
The experiments on AWS were conducted in the Northern Virginia (us-east-1) region from 5 am to 9 pm KST, spanning from September to December 2023. To minimize the impact of cloud user traffic and maintain consistent experimental conditions, tests were carried out across various availability zones (from us-east-1a to us-east-1f). For the GPU-VMs, we utilized $\mathtt{g3.xlarge}$(NVIDIA M60), $\mathtt{g4.xlarge}$(NVIDIA T4), and $\mathtt{g5.xlarge}$(NVIDIA A10G), and for the CPU-VMs, we used the CPU type instances specified in Table \ref{tbl:awstypes}. The instances’ storage was selected as AWS's EBS(gp2). Due to AWS's vCPU allocation policy, our usage of GPU-VMs was limited to a maximum of 32 out of 128 vCPUs.
To validate the efficiency of \tbdfixed{}, we employed major image processing models like ResNet-18 and EfficientNet-v2-m, using the CIFAR-10 dataset. The batch size for single GPU processing was fixed at 128, and each DDL training was performed up to 50 epochs utilizing the Adam optimizer with the $\mathtt{ReduceLROnPlateau}$ scheduler. We report an average of 5 runs for each experiment. 
{To complement our workload, we also considered using ResNet-18, ResNet-34, ResNet-50, ResNet-152, as well as EfficientNet-V2L models.}



We use two metrics to evaluate the performance of \tbdfixed{}.
\squishlist
\item
\textbf{Total cost:} 
The actual cost incurred in executing the workload and 
calculated based on the number of instances, the hourly rate, and the makespan. It is used to evaluate the cost-efficiency of a specific VM cluster configuration.
\item
\textbf{Makespan:}
The total time required to complete the workload execution.
The makespan duration impacts the overall cost, and is directly proportional to 
preemption probability of spotVM. 
It is used to assess how quickly training was conducted using a specific VM cluster configuration.
\squishend

\subsection{Evaluation of Overhead Modeling}
\label{sec:overhead_expr}
To accurately estimate the overhead that arises when increasing the number of GPU-VMs, \tbdfixed{} introduces a scaling factor. When employing a tiering structure, it predicts the point where network transmissions become saturated and prevents the network transmission bottlenecks. 
We verify that the overhead modeling in \tbdfixed{} is accurate and applied effectively.

\subsubsection{Scaling Factor for Multiple GPUs}
\tbdfixed{} 
estimates the speedup when using $n$ VMs. 
To fairly validate its prediction, we trained the ResNet50, a workload not used in regression, for 30 epochs 5 times for each GPU-VM. 
Then, we compared the empirically measured speedup derived from the average execution time and the speedup estimated by \tbdfixed{}.

\begin{figure}[!tb]
\centering
    \begin{tabular}{@{}c@{}c@{}}
        \includegraphics[width=0.5\textwidth]{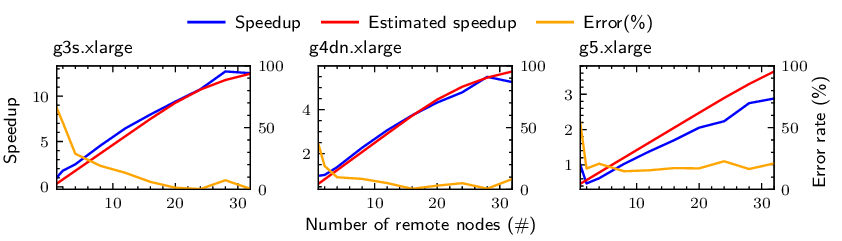}
        \vspace{-5pt}
    \end{tabular}
    \caption{Overhead analysis 
    in modeling. 
    Results show the estimated speedup (red) and actual speedup (blue) when training ResNet50 for 30 epochs, as the number of GPU-VMs increases for three different GPU-VMs. 
    }
    \vspace{-10pt}
\label{fig:scaling_eval}
\end{figure}
\index{figure}

Figure \ref{fig:scaling_eval} presents the results. 
The actual measured speedup and estimated speedup for all GPU-VMs showed similar tendencies. 
The error represents the error rate between the actual and estimated speedups. 
The error reached up to 66.3\% for $\mathtt{g3s.xlarge}$ when the node count was low. 
However, it exhibited an error rate below 10\% from 16 nodes for $\mathtt{g3s.xlarge}$ and from 4 nodes for $\mathtt{g4dn.xlarge}$. 
This demonstrates that as the number of GPU-VMs in the cluster increased, \tbdfixed{}'s prediction became more accurate.

On the other hand, for $\mathtt{g5.xlarge}$, the error rate remained around 20\% regardless of the increase in node numbers. 
This is because the nature of $\mathtt{g5.xlarge}$ means that as distributed learning is conducted across more clusters, the influence of its overhead increases, making the actual speedup lower than the estimated speedup by \tbdfixed{}. To better understand these results, additional experimentation with g5.xlarge would be beneficial.
However, given the similar speedup trend, we can conclude that \tbdfixed{} provides reasonably accurate predictions.

\subsubsection{Network Bandwidth Saturation Point}

{In the tiering architecture, network bandwidth can become a bottleneck if too many checkpoint shards are transferred to a single CPU-VM (On-Demand VM), which primarily serves as a remote fat `persistent' memory server. 
We aimed to validate the effectiveness of the network saturation table (referenced as Table \ref{tab:n_sat}) used by \tbdfixed{}. This table considers the network bandwidths of both GPU-VMs and CPU-VMs and recommends an appropriate number of CPU-VMs based on these considerations.

{To verify the validity of these recommendations, we conducted experiments. Specifically, \tbdfixed{} suggested using 32 Spot VMs of $\mathtt{g4dn.xlarge}$ and 3 On-Demand VMs of $\mathtt{c4.xlarge}$, considering a pricing willingness of \$6.5. This recommendation was evaluated by measuring the transmission time while progressively increasing the number of CPU-VMs from 1 to the maximum of 7, as allowed by the pricing willingness (note that 32 is the maximum number of VMs selectable under AWS policies)}.




\begin{figure}[!t]
\centering
    \begin{tabular}{@{}c@{}c@{}}
        \includegraphics[width=0.5\textwidth]{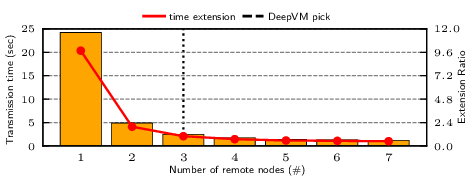}
        \vspace{-5pt}
    \end{tabular}
    \caption{Network transmission time analysis. 
    When PW is set to \$6.5, we measured the transmission time by varying the number of CPU-VMs in the configuration recommended by \tbdfixed{}. 
    Note that \tbdfixed{} recommendes 3 remote nodes (VMs).
    From 3 VMs onwards, the time doesn't significantly decrease, but if it drops below 3 VMs, the transmission time sharply increases.
    {The orange bars (histogram) represent the network transmission time, while the red line indicates the ratio of this time compared to the transmission time picked by \tbdfixed{}.
    In the plot, the remote node is the On-Demand VM.
    }     
    }
    \vspace{-23pt}
\label{fig:saturation_eval}
\end{figure}
\index{figure}

{Figure \ref{fig:saturation_eval} illustrates the results. This comparison includes the transmission time recommended by \tbdfixed{} (3 VMs), as well as the transmission times observed with varying numbers of CPU-VMs. Notably, when only one On-Demand VM was utilized, the transmission time was nearly 10 times higher compared to the \tbdfixed{} recommendation. 
However, at the other end of the spectrum, with 7 On-Demand VMs, the transmission time was 51.58\% lower than the recommended time (which is the transmission time when 3 remote nodes are used}. 
These findings indicate that the recommendations made by \tbdfixed{} are reasonable and effectively reflect network saturation.
}


\subsection{Effectiveness of \tbdfixed{}{}}

\subsubsection{Baseline}
In this experimental series, we evaluate \tbdfixed{}'s configuration policy against various other policies to verify how effectively \tbdfixed{} can configure a range of instances. As mentioned in Section \ref{sec:limitation}, no existing studies were found that were conducted in a similar environment. Therefore, for baseline comparison, we introduce three distinct policies (algorithms).
All algorithms take into account the user Pricing Willingness (PW). In the simulation, instances can be utilized without a limit on their number as long as their hourly cost does not exceed PW. In actual experimental environments, however, the maximum number of instances is determined by AWS policies.
\squishlist
\item
Cost-First: This policy utilizes the most cost-effective GPU-VM instances per hour as the single anchor. This setup aims to use as many instances as possible.
\item
{Performance-First: This policy is oriented towards GPU-VM instances that have the highest FLOPS, often correlating with higher costs and potentially fewer instances used.}
\item
\tbdfixed{}-Noscale: This policy employs the same algorithm as \tbdfixed{}, but does not consider the $ScalingFactor$. Consequently, it may select instances with high FLOPP but less suitable for parallel processing.
\squishend

{Cost-First and Performance-First focus primarily on the GPU-VM aspect, thus both algorithms essentially consider a Single-Anchor architecture in their approach.}

\subsubsection{Validation in simulation}
\begin{figure}[!tb]
\centering
    \begin{tabular}{@{}c@{}c@{}}
        \includegraphics[width=0.48\textwidth]{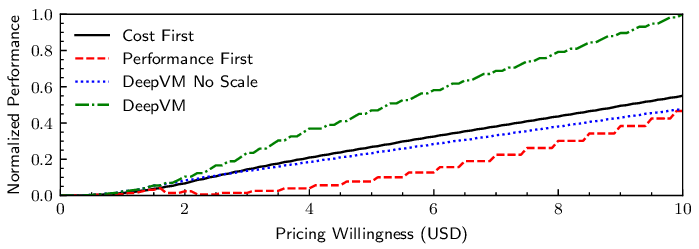}
        \vspace{-5pt}
    \end{tabular}
    \caption{This graph represents the normalized performance of VM cluster configurations recommended by Cost-First, Performance-First, \tbdfixed{}-NoScale, and \tbdfixed{} policies across different levels of Pricing Willingness (PW) ranging from \$0 to \$10. Performance is normalized to the highest performance observed for \tbdfixed{} at PW=\$10.}
    \vspace{-10pt}
\label{fig:simulation_result}
\end{figure}
\index{figure}
The simulation was conducted by varying the Pricing Willingness (PW) from \$0 to \$10 in increments of \$0.1. For each PW value, the recommended configurations from four policies (Cost-First, Performance-First, \tbdfixed{}-Noscale, and \tbdfixed{}) were evaluated. Performance was assessed for each configuration based on the number of GPU-VMs, eFLOPS, and a performance function utilizing the $ScalingFactor$. The results of these experiments are depicted in Figure~\ref{fig:simulation_result}.

Our findings demonstrate that \tbdfixed{} consistently delivered the highest performance across all PW values. In contrast, \tbdfixed{}-Noscale showed poorer performance than Cost-First when PW exceeded \$2.5, attributed to the presence of instances with high FLOPP but subpar parallel processing capabilities.
The Cost-First policy, despite increasing PW values, did not show significant performance improvements compared to \tbdfixed{}. This is because it only considers cheaper instances, thus failing to account for instances that offer better performance relative to their cost.
Performance-First experienced a sharp decrease in performance starting at a PW of \$1.6. This decline occurred because the policy focuses solely on performance, missing the opportunity to utilize multiple instances. However, as PW increases, Performance-First gradually improves in performance by considering multiple high-performing instances.

{Interestingly, for all four policies - Cost-First, Performance-First, \tbdfixed{}-Noscale, and \tbdfixed{} - the performance improvement follows a stepwise pattern beyond certain PW thresholds. This phenomenon occurs because, for each policy, the optimal instance selection stabilizes beyond a specific PW value. As a result, every time the PW increases enough to include an additional instance of this optimal type, there is a noticeable incremental improvement in performance. 
This consistent pattern across different policies highlights how performance scales in a stair-step manner as PW allows for the inclusion of more instances of the most efficient type as per the respective policy's criteria. We observed that this trend consistently persists even beyond a PW of \$10.}



In conclusion, the three baseline policies fail to provide the optimal combination of instances. However, \tbdfixed{} effectively recommends the best possible combination by comprehensively considering various factors.

\subsubsection{Validation in AWS}

To validate the effectiveness of the VM cluster configuration proposed by \tbdfixed{} in a real AWS environment, we created instances on AWS and executed DDL workloads. Concurrently, VM cluster configurations suggested by three baseline policies were also included in the experiment, considering the top 3 configurations of \tbdfixed{} and one configuration according to each baseline policy. This comparative analysis was conducted to assess the cost efficiency of \tbdfixed{}.
For a more realistic setup, the pricing willingness input for \tbdfixed{} was set to \$3, based on the spot price of \$2.5248 for 16 $\mathtt{g4dn.xlarge}$ instances. The experiments applied training workloads using the ResNet50 model.

\begin{figure}[!t]
\centering
    \begin{tabular}{@{}c@{}c@{}}
        \includegraphics[width=0.48\textwidth]{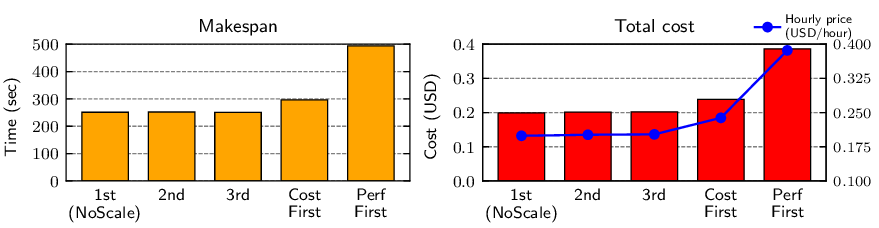}
        \vspace{-5pt}
    \end{tabular}
    \caption{Executing the same workload with PW=\$3, Buffer size=2, and checkpoint file size=300MB, comparing \tbdfixed{} configuration (1st, 2nd, 3rd) and baseline policies (Cost-First, Perf-First, NoScale).}
    \vspace{-10pt}
\label{fig:aws_eval}
\end{figure}

The experimental results are depicted in Table \ref{tbl:rank} and Figure~\ref{fig:aws_eval}. Table \ref{tbl:rank} shows the VM cluster configurations as recommended by \tbdfixed{}, where the first recommendation is the least costly and the sequence follows an increasing order of total cost. This pattern demonstrates that \tbdfixed{} recommends configurations that effectively balance cost and performance. In contrast, the configurations suggested by the Cost-First and Performance-First policies led to higher total costs than those recommended by \tbdfixed{}.
The \tbdfixed{}-NoScale policy recommended a configuration identical to the first configuration of \tbdfixed{}. This outcome is due to the absence of instances with high FLOPP but poor parallel processing capabilities among the provided instances.

Figure~\ref{fig:aws_eval} compares the makespan, total cost, and hourly price of each configuration. Across all scenarios, \tbdfixed{}’s recommendations resulted in the shortest makespans. The configurations suggested by the Cost-First and Performance-First policies, however, exhibited longer makespans, leading to higher overall costs. Particularly, the Performance-First policy's configuration nearly doubled the makespan and total cost compared to \tbdfixed{}’s recommendations.

\begin{table}[!t] 
    \begin{center}
    \caption{\small Cost and Configuration Comparison for 30-Epoch ResNet50 Training at PW=\$3: Top three configurations from \tbdfixed{} and bottom three from baseline algorithms.}
    \label{tbl:rank}
    \resizebox{0.48\textwidth}{!}{
        \begin{tabular}{|c|c|c|c|c|c|}
        \hline
        Configuration & Total cost (USD) & GPU & \# of GPU & CPU & \# of CPU \\ \hline \hline
        \tbdfixed{} 1st & 0.1991 & g4dn.xlarge & 17 & c5.xlarge & 1 \\ \hline
        \tbdfixed{} 2nd & 0.2013 & g4dn.xlarge & 17 & m6i.xlarge & 1 \\ \hline
        \tbdfixed{} 3rd & 0.2022 & g4dn.xlarge & 17 & c5n.xlarge & 1 \\ \hline
        \hline
        \tbdfixed{}-NoScale & 0.1991 & g4dn.xlarge & 17 & c5.xlarge & 1 \\ \hline
        Cost-First & 0.2385 & g4dn.xlarge & 16 & - & - \\ \hline
        Performance-First & 0.3865 & g5.xlarge & 7 & - & - \\ \hline
    \end{tabular}
    }
    \end{center}
    \vspace{-14pt}
\end{table}

\index{figure}

From these results, two key facts can be inferred:
(i) \tbdfixed{} does not merely recommend configurations with low hourly prices. Instead, it provides recommendations based on a comprehensive consideration of various cloud VM's variables.
(ii) Recommendations by \tbdfixed{} do not compromise makespan delay just for a lower price.

\subsection{Portability of \tbdfixed{}}
{In the current state, \tbdfixed{} is limited to only certain AWS instances. For utilization on new instances or different cloud services, instance-specific information must be pre-programmed into the algorithm. This data can be readily updated by open-source users or corresponding cloud service provider (CSP) administrators. Once updated, this information enables other users to effortlessly apply \tbdfixed{} to these instances.}
\section{Conclusion}
\label{sec:conc}
In this paper, we introduced \tbdfixed{}, a strategy for efficiently and economically utilizing Spot and On-Demand VMs in distributed deep learning, with a focus on overcoming the checkpointing challenges in Spot VM environments.
Our simulations and AWS deployments demonstrated that \tbdfixed{} significantly reduces costs while maintaining high performance, outperforming existing baseline models.
{`\tbdfixed{}` is designed for flexibility and adaptability to various AWS instances and cloud services, requiring only updates to instance-specific data. Though focused on an image processing deep learning model, the method is broadly applicable to different deep learning workloads. This is achieved by customizing benchmarking metrics like `eFLOPS' and overhead functions to suit specific deep learning tasks and instances, ensuring its effectiveness across diverse scenarios.}

\section*{Acknowledgement}
{This work was supported by the Institute of Information \& communications Technology Planning \& Evaluation (IITP) grant funded by the Korea government (MSIT) (No.2022-0-00498, Development of high-efficiency AI computing SW core technology for high-speed processing of large learning models). This work was also supported by, and used the resources of, the Oak Ridge Leadership Computing Facility, located in the National Center for Computational Sciences at ORNL, which is managed by UT Battelle, LLC for the U.S. DOE (under the contract No. DE-AC05-00OR22725).}

{
\setstretch{0.99}
\bibliographystyle{ieeetr}
\bibliography{paper}
}

\setstretch{0.98}
\clearpage
\begin{appendices}
\section{Artifact Description}
\label{sec:artifact}

\subsection{Abstract}
In this document, we provide detailed instructions and steps that enable readers to:
(i) install all programs, source codes, and dependencies related to the LP-based \tbdfixed{};
(ii) reproduce experiments in the same environment to extract data;
(iii) generate figures from the data;
(iv) create scenarios for customized experiments.

\subsection{Description}

\subsubsection{Check-list (artifact meta information)}
\begin{itemize}
    \item
    \textbf{Program:}
    \begin{enumerate}
        \item \tbdfixed{} (LP modeling) and its simulation;
        \item Software and example codes for each architecture for effectiveness experiments;
        \item Training example codes for overhead modeling validation experiments;
    \end{enumerate}
    \item 
    \textbf{Dataset:}
    Information of available instances as of the experiment time, virtual instance information based on Amazon environment, workload (CIFAR-10);
    \item
    \textbf{Hardware:}
    Various x86 or x64 CPUs (for \tbdfixed{} and simulation), 
    AWS instances (for training)
    \item 
    \textbf{Experiment workflow:}
    \begin{enumerate}
        \item \tbdfixed{}
            \begin{enumerate}
            \item Run the software after installation.;
            \item If verification is desired, launch instances in the cloud as per recommended configuration;
            \item Measure the execution time and cloud cost to generate tables and figures;
            \end{enumerate}
        \item Simulation: Run the simulation after installation. Generate figures with the results;
        \item Overhead modeling validation: Run the training code after installation. Generate figures with the results;
    \end{enumerate}
    \item
    \textbf{Output:}
    Top n recommended configurations (\tbdfixed{}),
    Data and graphs showing the performance of the policy and the PW value (Simulation),
    Data and graphs representing the validation results
    \item
    \textbf{Experiment customization:} See below;
    \item
    \textbf{Publicly available?:} Yes.
\end{itemize}

\subsubsection{How the artifact can be obtained}
The Persistent ID of the artifact is:

\cb{https://doi.org/10.5281/zenodo.10672948}

\noindent Furthermore, to ensure that readers can use the same version of the software for experiment reproduction, the code and necessary files along with instructions have been archived in Software Heritage:

\cb{https://archive.softwareheritage.org/
swh:1:dir:561f421fa40bdc4fdb1f9f0c63b8f93e73886717;
origin=https://github.com/lass-lab/DeepVM}

\noindent Finally, all source materials can also be downloaded from the git repository associated with the Persistent ID:

\cb{https://github.com/lass-lab/DeepVM.git}

While the content is identical across all platforms, be aware that the git repository may be updated over time.

    
    

The artifact consists of four main directories:
\begin{itemize}
    \item
    $\mathtt{solution:}$ 
    Includes the LP modeling and simulation programs, as well as the instance dataset and dataset generation program used.
    
    \item
    $\mathtt{arch\_sw:}$ 
    Contains example source codes for distributed learning per architecture to reproduce the effectiveness experiments of \tbdfixed{} in the paper.
    
    \item
    $\mathtt{validation:}$ 
    Includes training source codes and usage instructions to reproduce the overhead modeling experiments in the paper.

    \item
    $\mathtt{results:}$
    Contains all source codes to generate the graphs presented in the paper using the result files.
    
\end{itemize}

\subsubsection{Hardware requirements}
\tbdfixed{} can be run on most x86 or x64 CPUs. However, using too many instance data may increase execution time depending on CPU performance. For validation or conducting experiments based on DeepVM results, it can be performed in a cloud environment. It is necessary to be able to launch instances with GPUs as both spot and ondemand types. Each instance must be able to communicate with each other, and be launched without incurring additional network communication costs depending on the region. This paper conducted experiments in an AWS environment, a representative CSP, and readers wishing to replicate the experiment in the same environment can secure sufficient quotas by consulting AWS.

\subsubsection{Software requirements}
\begin{itemize}
    \item
    Git:
    A version control system storing all materials for simulation codes, execution codes, datasets, and extracting result files.
    \item
    Python 3 (version 3.10.6):
    Used for LP modeling and execution, data extraction, and graph generation. Also used in architecture-specific software for deep learning. Required libraries and their versions include: numpy(1.25.0), pandas(2.0.2), torch(2.0.1), torchvision(0.15.2), matplotlib(3.7.1), jupyter nodebook(6.5.4)
    \item
    pip (version 22.0.2):
    Used to download and manage libraries needed for running Python 3.
    \item 
    OpenMPI (version 4.1.4):
    Used for data transmission in multi-level checkpointing when using software for Tiering architecture.
    \item 
    GCC (version 11.4.0):
    Used to compile modules implemented in C++ in the Tiering architecture.
\end{itemize}
Readers who do not wish to use Git can download the artifact from Software Heritage as mentioned earlier. The required Python Packages are stored in $\mathtt{requirements.txt}$ in each directory.

\subsubsection{Datasets}
There are two types of datasets for modeling: virtual instance data and real instance data. Virtual instance data simulates AWS environment instances, and real instance data consists of information about instances available at the time of the experiment. Network saturation point data is embedded in $\mathtt{.py}$ files, and the scaling factor of actual instances is built into the \tbdfixed{} source code. All other data is stored in $\mathtt{.json}$ format. Detailed content of the dataset is described in the experiment customization section.
For training, the CIFAR-10 dataset was used. This dataset is automatically downloaded to the local computer at the start of training by PyTorch, and can also be downloaded using the provided shell file.

\subsubsection{Installation}
First, download the artifact (using Software Heritage or Git). If using Git, the following command can be used to copy the Git repository to the local computer: \texttt{git\ clone\ https://github.com/lass-lab/DeepVM.git}. If necessary, GCC, Python 3, pip, OpenMPI should be installed additionally. For packages mentioned in the software requirement section, they can be installed directly or using the \texttt{pip\ install\ -r\ requirements.txt} command.

\subsubsection{Experiment Workflow}
\begin{itemize}
    \item
    \textbf{\tbdfixed{}:}
    To run \tbdfixed{}, navigate to the \texttt{./solution} directory and execute the command: \texttt{./run.py}. 
    Parameters including PW can be entered along with the command. If run without entering any parameters, default parameters are automatically entered: \texttt{pw=3, bf=2, ckpsize=0.5, MAX\_LIMIT=256}.
    For detailed usage including parameter input, refer to $\mathtt{DeepVM/solution/README.md}$.
    
    \item
    \textbf{Simulation:}
    To perform the simulation, navigate to the \texttt{./solution} directory and execute the command: \texttt{./simulation.py}.
    The usage of parameters is similar to \tbdfixed{}; refer to the same README file for guidance.
    When executed, a \texttt{.pkl} file containing the simulation results will be generated.

    \item
    \textbf{Validation (Effectiveness):}
    Readers who want to validate a specific configuration can do so by creating the same cluster on AWS.
    Distributed learning is performed using the software suitable for the recommended architecture.
    To prepare for distributed learning, copy all files in the respective directory to the user's home directory, then prepare with the following command: \texttt{bash ./initialization.sh}.
    After preparation, distributed learning can be started with the command: \texttt{bash ./run\_experiments.sh}.
    For detailed execution methods, refer to the architecture-specific README in the $\mathtt{arch\_sw}$ subdirectory.
    
    \item
    \textbf{Validation (Scaling Factor):}
    After creating the maximum number of identical instances, copy all files under $\mathtt{validation/scaling\_factor}$ to the master's home directory.
    Then, after preparing similarly to the single anchor architecture software,
    perform distributed tasks with the command: \texttt{bash ./run\_experiments}.
    This shell automatically records the time while increasing the number of training nodes.
    For detailed execution methods and examples, refer to the README in the $\mathtt{validation/scaling\_factor}$ subdirectory.
    
    \item
    \textbf{Validation (Network Saturation):}
    First, create the desired cluster on AWS.
    While maintaining the number of GPU spot instances in the cluster, repeat the experiment by increasing the number of CPU spot instances.
    The method of preparation and execution for distributed learning is similar to the previously described tiering architecture software.
    For detailed execution methods and examples, refer to the README in the $\mathtt{validation/network\_saturation}$ subdirectory.
    
\end{itemize}

\subsubsection{Evaluation and Expected Result}
\begin{itemize}
    \item
    \textbf{\tbdfixed{}:}
    The four policies mentioned in the experiment section of the paper will output their respective best configurations. However, for \tbdfixed{} policy, only the top 3 are outputted. Readers can launch clusters with these configurations to validate their effectiveness.

    \item
    \textbf{Simulation:}
    A $\mathtt{simul\_results.pkl}$ file is created. This file contains the performance of configurations proposed by the four policies for each PW value in dictionary form. To generate graphs using this file, use the following Jupyter notebook file:
    
    $\mathtt{results/simulation/graph.ipynb}$

    \item
    \textbf{Validation (Effectiveness):}
    A $\mathtt{.result}$ file is generated per experiment. This file can be opened with a text editor and contains the \texttt{makespan} time consumed in training.
    Readers can i) collect the makespan from various $\mathtt{.result}$ files,
    ii) collect the hourly price of each configuration,
    iii) enter these into the $\mathtt{result/real\_world}$ directory's $\mathtt{fig8\_data.xlsx}$ file.
    To generate graphs using this file, use the following notebook file:

    $\mathtt{results/real\_world/graph.ipynb}$
    
    \item
    \textbf{Validation (Scaling Factor):}
    Once training is completed, a file with all recorded data is created in the master node's home directory.
    This data includes the number of training nodes and the training time logged for each.
    Record this in the appropriate column of $\mathtt{results/fig5\_data.xlsx}$.
    To generate graphs using this file, use the notebook file mentioned in effectiveness.

    \item
    \textbf{Validation (Network saturation):}
    Each checkpointing yields \texttt{[serialization]} and \texttt{[sending]} logs. The number of logs per checkpointing can vary depending on the number of shards and remote nodes. The average time difference between serialization and sending pairs with the same sharding-checkpointing number is recorded as the time for that checkpointing.
    Average several checkpointing times to record the transmission time of that cluster in $\mathtt{results/fig6\_data.xlsx}$.
    To generate graphs using this file, use the notebook file mentioned in effectiveness.
\end{itemize}

\subsubsection{Experiment customization}
This section describes how to run user-defined scenarios using this artifact.
Examples of user-defined scenarios include changing the type of virtual or real instances, changing prices, modifying performance figures, and altering the number of bottlenecks per network bandwidth.
To change the type, price, or metrics of instances, modify the $\mathtt{.json}$ files in the directory. Then, \tbdfixed{} or simulation can be performed using the modified file. The values in the files consist of instance names, types, prices, parameters of regression equations, network performance, and availability, and detailed information can be found in the artifact's README. The method of creating figures using the result data is the same as in the original scenarios, but some scenarios may require additional code changes.

\end{appendices}

\end{document}